\documentclass[prl,aps,twocolumn]{revtex4}
\usepackage{graphicx} 
\usepackage[usenames]{color}
\usepackage{amsmath,amssymb}
\usepackage{gensymb}
\usepackage{natbib}
\usepackage{xcolor}
\def\strutdepth{\dp\strutbox}
\def\nw#1{\strut\vadjust{\kern-\strutdepth\vtop to0pt{\vss\hbox to\hsize
{\hskip\hsize\hskip5pt$\leftarrow$\hss\strut}}}{\em #1}}
\usepackage{rotating}

\begin{document}

\title{Dynamical compressibility of dense granular shear flows}
\author{Martin Trulsson} \author{Mehdi Bouzid }  \author {Philippe Claudin} \author{Bruno Andreotti}
\affiliation{Physique et M\'ecanique des Milieux H\'et\'erog\`enes, UMR 7636 ESPCI -- CNRS -- Univ.~Paris-Diderot -- Univ.~P.M.~Curie, 10 rue Vauquelin, 75005 Paris, France}
\begin{abstract}
It has been conjectured by Bagnold \cite{Bagnold:66} that an assembly of hard non-deformable spheres could behave as a compressible medium when slowly sheared, as the average density of such a system effectively depends on the confining pressure. Here we use discrete element simulations to show the existence of transverse and sagittal waves associated to this dynamical compressibility. For this purpose, we study the resonance of these waves in a linear Couette cell and compare the results with those predicted from a continuum local constitutive relation.
\end{abstract}

 \date{\today}

\maketitle

Acoustics in granular media is a rapidly developing field that is attractive both from the fundamental point of view and for its applications in material science, soil mechanics and geophysics \cite{TG10,A12}. It provides a unique tool to probe the mechanical response of such materials, to detect heterogeneities, aging or avalanche precursors.  In the \emph{static} case, grain elasticity is the only restoring force; elastic waves present most of the known behaviors of mechanical waves: velocity dispersion, non-linearity associated to contact geometry, scattering associated to disorder, anisotropy, aging, mode coupling, wave-guiding in the presence of macroscopic heterogeneities, fragility, etc \cite{Goddard:90,JMS05,JP05,DNHJ06,TG09,TG10}.  Recent studies  \cite{MGJS99,MGJS04,BAC08} have focused on the failure of the mean field description of elasticity when approaching the jamming transition separating the solid-like from the liquid-like regime. The importance of non-affine displacements in this limit was revealed by the structure of the vibration spectrum, which exhibits an anomalous excess of low frequency `soft' modes  \cite{Wyart:05,vH10,Gomez:12}. In the following we wish instead to investigate wave propagation on the \emph{liquid} side of the jamming transition.

Wave propagation in granular shear flows have never been studied for itself. It has been revealed indirectly by instabilities, in situations where acoustic waves are spontaneously emitted during a dense granular flow (see \cite{A12} for a review): when sufficiently dry, avalanches flowing at the surface of a dune produce a loud sound known as the `song of dunes' \cite{A04,DMHEPDK06,AB09,Staron:12}; vibrations are also produced during the discharge of a granular flow from a silo or any elongated tube \cite{BCJA10}; soft mono-disperse particles flowing on an inclined plane near the angle of repose exhibit spontaneous oscillations at the resonant frequency of elastic waves \cite{Silbert:05,Richard:12}. These phenomena have mostly been associated to the elastic deformations of the grains. However, Bagnold \cite{Bagnold:66} and followers \cite{DMHEPDK06,A12,Staron:12} have hypothesized the existence of vibrations driven by the coupling between normal stress and shear rate. 

In this Letter we theoretically demonstrate the existence of such waves in dense slow shear flows of \emph{hard non-deformable grains}. We show that these waves are due to a compressibility of dynamical origin, whose scaling with the volume fraction $\phi$ is related to the non-affine cooperative motion of the grains. The dispersion relation presents three branches, two of which becoming non-dissipative when approaching the jamming point. Our analytical predictions are confirmed by numerical simulations performed with quasi-rigid grains, which show no dependence on their elasticity nor on their restitution coefficient.
\begin{figure}[t!]
\includegraphics{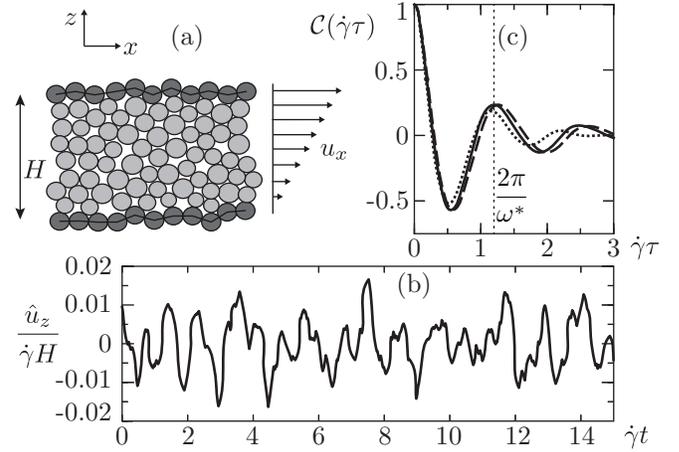}
\vspace{-5 mm}
\caption{(a) Schematic of the numerical set-up. (b) Typical velocity signal as a function of time $t$, in a simple shear flow at $\phi \simeq 0.78$, $H/d \simeq 40$, and $\dot \gamma d \simeq 0.1 \sqrt{{\mathcal P}/\rho}$. To improve the signal-to-noise ratio, we have computed the projection $\hat{u}_z$ of the vertical velocity field on the mode $\sin(\pi z/H)$.  (c) Autocorrelation functions $\mathcal{C}(\dot \gamma \tau)$ of the signal $\hat{u}_z(t)$, for different values of the parameters $\kappa_n$, $\kappa_t$ and $e$. The first maximum gives the spontaneous frequency $\omega^*/ 2 \pi$.}
\vspace{-5 mm}
\label{fig:signal}
\end{figure}
\begin{figure}[t!]
\includegraphics{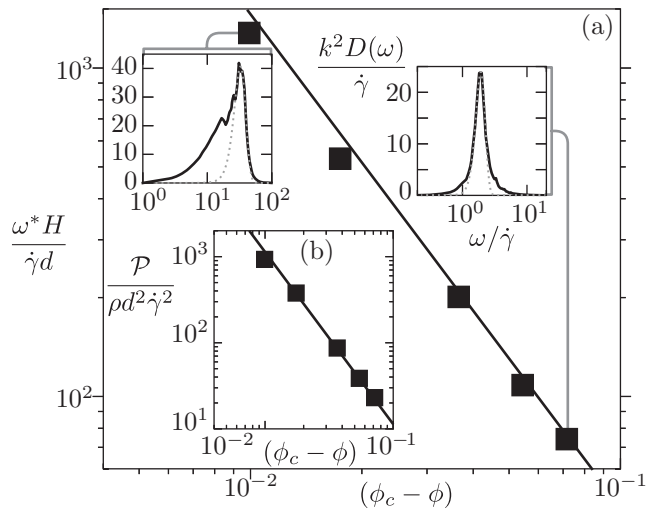}
\vspace{-5 mm}
\caption{(a) Spontaneous angular frequency $\omega^*$ as a function of $(\phi_c-\phi)$. Solid line: prediction for the resonant frequency $2\pi c/H$ deduced from Eq.~\ref{cc}. \textit{Insets:} Black solid line: power spectral density $D(\omega)$ of the projected vertical velocity $\hat{u}_z$. Gray dotted line: best fit by a gaussian. (b) Scaling of the pressure with the volume fraction. Solid line: best fit by the form $f(\phi)$ given by Eq.~(\ref{Eq:ff}).}
\vspace{-5 mm}
\label{fig:omegastar}
\end{figure}
\begin{figure*}[t!]
\includegraphics{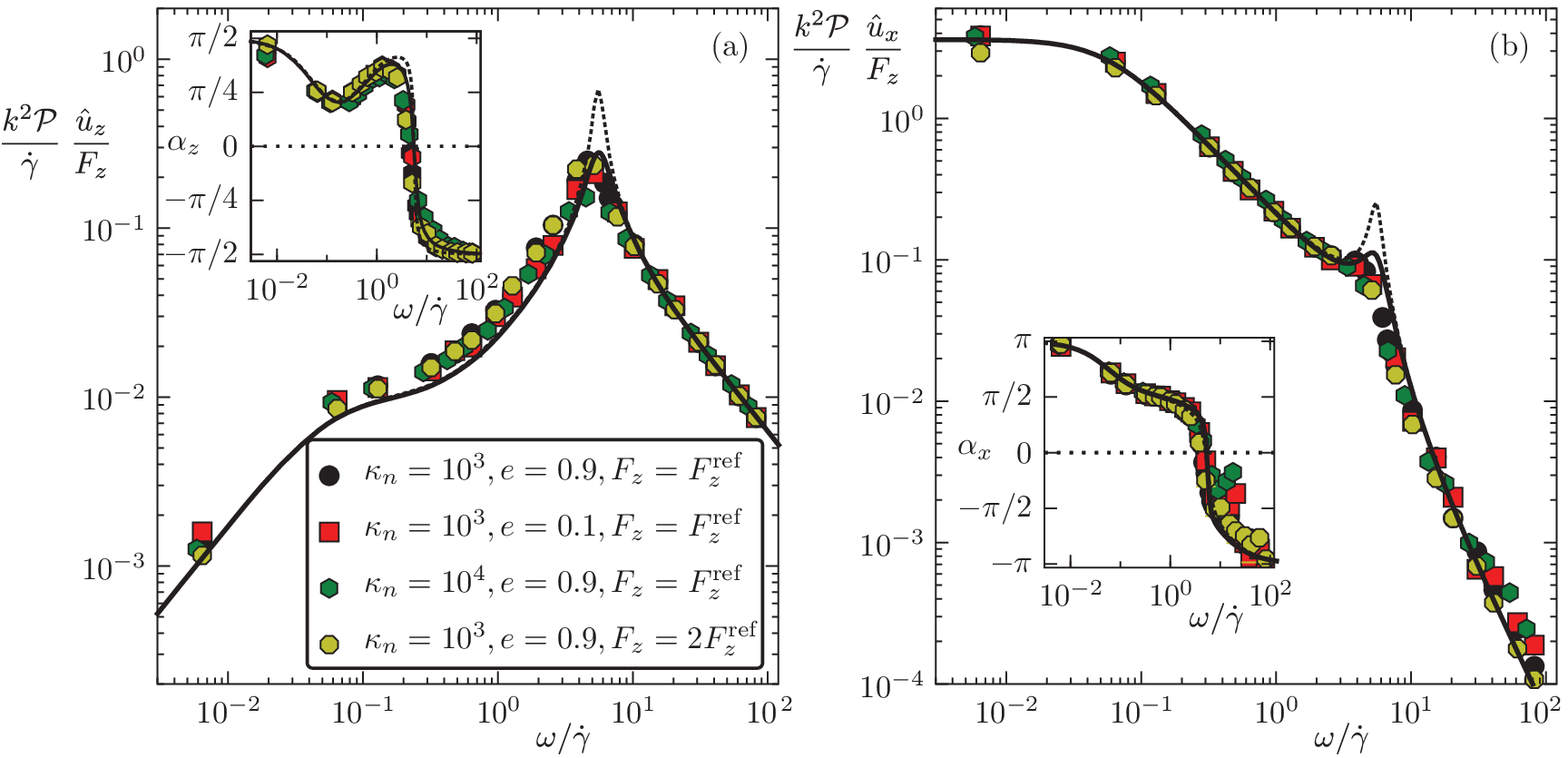}
\vspace{-5 mm}
\caption{(Color online)  Amplitude and phase of the velocity response along the $z$ axis (a) and the $x$ axis (b). Solutions of Eqs.~(\ref{Navier}), with forcing, are displayed in solid (with a bulk viscosity $g=2$) and dotted lines (for $g=0$). The velocity field is projected onto the form $\hat{u}_i \sin(kz) \sin(\omega t+\alpha_i)$, where $\hat{u}_i$ is the amplitude and $\alpha_i$ the phase shift with respect to the forcing. The various symbols correspond to different restitution coefficients, different spring constants and different amplitudes of the forcing. The simulation conditions are the same as in Fig.~\ref{fig:signal}, with $F_z^{\rm ref} \simeq \left( \frac{4 \phi}{\pi d^2}\right) f_{z}^{\rm ref} \simeq 0.025\, \mathcal{P}/d$.}
\label{fig:forcing_z}
\vspace{-5 mm}
\end{figure*}

{\it Numerical model~--~}We consider a two-dimensional system constituted of  $\simeq 2\cdot 10^3$ circular particles of mass $m_i$ and diameter $d_i$, with a $\pm$50 \% polydispersity (Fig.~\ref{fig:signal}a). The shear cell is composed of two rough walls distant by  $H$ and moving along the $x$-direction at a constant velocity. Simulations are therefore performed at a constant, imposed volume fraction $\phi$. Periodic boundary conditions are applied along $x$.  The particle dynamics is integrated using the Verlet algorithm. The particles are submitted to contact forces modeled as viscoelastic forces along the normal direction and as a Coulomb friction along the tangential direction \cite{Cundall79,daCruz05,Luding06}. Quantities used in the model are expressed in terms of the grain density $\rho$, of the average pressure $\mathcal{P}$, and of the mean grain diameter $d$. In this system of units, the normal spring constant $\kappa_n$ is chosen sufficiently large (between $10^3$ and $10^4$) to reach the rigid asymptotic regime. The Coulomb friction coefficient is chosen equal to $\mu_p=0.4$ and the tangential spring constant $\kappa_t =0.5 \kappa_n$. The influence of viscoelastic parameters is encoded in the restitution coefficient $e$, which is varied between $0.1$ and $0.9$. The average shear rate $\dot \gamma$ is kept constant during all simulations.

{\it Spontaneous and forced oscillations~--~}Looking at the time fluctuations of the velocity field (Fig.~\ref{fig:signal}b),  simple shear flows are found to present spontaneous vibrations. As seen from the velocity autocorrelation function (Fig.~\ref{fig:signal}c) and the power spectral density (Fig.~\ref{fig:omegastar}) of the velocity signal, one observes that this vibration is not a broadband noise but oscillations at a well defined angular frequency $\omega^*$. Keeping the shear rate $\dot \gamma$ constant and approaching the jamming point $\phi_c$, $\omega^*$ is found to diverge as $(\phi_c-\phi)^{-3/2}$ (Fig.~\ref{fig:omegastar}a). The coherence time can be determined from the spectrum, which presents a narrow peak whose width at half maximum is around $\omega^*/4$. (Fig.~\ref{fig:omegastar}).  Furthermore, as $\phi\to\phi_c$, it develops a tail towards small frequencies, which is reminiscent of the spectral density of elastic modes on the solid side of jamming, for $\phi>\phi_c$.


To analyze the vibration mode by mode, we have performed simulations in which each grain is submitted to a force along $z$ of the form $f_{z} \sin(\omega t)\sin(kz)$, with $k=\pi/H$. We analyze the response of the system to this forcing with a simple `lock-in amplifier': we compute the Fourier component of the velocity field at the angular frequency $\omega$ and the wavenumber $k$. The amplitude of the forcing is chosen small enough to remain in the linear regime. The velocity response is then proportional to the forcing $f_z$ and presents a resonance, characterized by a maximum of amplitude and by a vanishing phase (Fig.~\ref{fig:forcing_z}). The resonant angular frequency coincides with the angular frequency $\omega^*$ of spontaneous oscillations (Fig.~\ref{fig:signal}).

The resonant curves  (Fig.~\ref{fig:forcing_z}) and the autocorrelation function of spontaneous vibrations (Fig.~\ref{fig:signal}c) turn out to be insensitive to the values of the restitution coefficient $e$ and of the spring constants. The lack of dependence on $e$ shows that the dense flow does not behave like a granular gas \cite{Savage:88,Goldshtein:96,Bougie:02,Amarouchene:06}, whose compressibility is controlled by binary collisions. The nature of these oscillations is also fundamentally different from those exhibited in \cite{Silbert:05,Richard:12} which are specifically observed for mono-disperse grains and whose frequency scales as $\sqrt{\kappa_n}$. The  vibration modes of a system of hard sheared particles are therefore different from both elastic modes in solids and sound waves in granular gases.

{\it Continuum model~--~} To shed light on the origin of these spontaneous oscillations, we consider a description of granular material as a continuum compressible medium characterized by the volume fraction  $\phi$ and the velocity field $\mathbf{u}$. Under the assumption that the rheology is local \cite{GDRMidi,JFP06}, dimensional analysis constrains the form of the constitutive relation between the stress  $\sigma_{ij}$ and the strain rate ${\dot \gamma}_{ij} = (\partial_i u_j+\partial_j u_i)$ to:
\begin{eqnarray}
\frac{\sigma_{ij}}{\rho d^2}&=& f(\phi)  |\dot \gamma|  \left[  -|\dot \gamma| \,\delta_{ij} + \mu(\phi){\dot \gamma}_{ij} + g(\phi) \sum_l {\dot \gamma}_{ll} \,\delta_{ij} \right] \label{rheononlocal}
\end{eqnarray}
where $\mu(\phi)$ is the effective friction coefficient, $f(\phi)$ the rescaled dilatancy, and $g(\phi)$ is the rescaled bulk viscosity.  As shown in Fig.~\ref{fig:omegastar}b, $f(\phi)$ diverges as 
\begin{equation}
f(\phi) = \frac{b^2}{(\phi_c-\phi)^2} 
\label{Eq:ff}
\end{equation}
at the critical volume fraction $\phi_c$ \cite{Boyer:11,Lerner:12,Andreotti:12,Trulsson:12}. The friction is regular in the limit $\phi\to \phi_c$ and is well approximated by 
\begin{equation}
\mu(\phi) = \mu_c + \frac{\delta \mu}{1+a b / (\phi_c-\phi)}.
\label{Eq:mu}
\end{equation}
We hypothesize  that the rescaled bulk viscosity is regular as well: $g\to g_c$ as $\phi\to \phi_c$. In a previous study \cite{Trulsson:12}, we have obtained $\phi_c \simeq 0.817$, $a=0.36$, $b=0.33$ and $\delta \mu=0.57$ in the 2D case, using homogeneous shear flows. $g$ has never been measured so far. Importantly, the equations are the same in both 2D and 3D.

The compressible Navier-Stokes equations are given by 
\begin{equation}
\rho \phi \frac{D \mathbf{u}}{Dt} = \nabla \boldsymbol{\sigma}+\mathbf{F} \quad{\rm and}\quad\frac{D \phi}{Dt}=-\phi (\nabla \cdot \mathbf{u}),
\end{equation}
where $\mathbf{F}$ is a bulk force. The base state is a homogeneous steady-state shear flow, and from now on we denote its volume fraction by $\phi$, its shear rate by $\dot \gamma$ and its pressure by $\mathcal{P}=f(\phi)\rho d^2  |\dot \gamma|^2$. Linear perturbations in volume fraction and velocities around the base state will be denoted by $\phi^1$ and $u^1_{i}$. Plane waves can only propagate along the  $y$ and $z$ axis as a plane wave emitted along $x$ would rotate due to convection by the base velocity field $u_x= \dot \gamma z$. This effect can be alternatively seen as a refraction by an index gradient.  

Let us consider first the case of waves propagating along the transverse axis (in 3D), of the form $\exp(i\omega t+iky)$. In order to induce compression, the wave must have a velocity component $u_y^1$ along $y$, which corresponds to a strain perturbation $\dot \gamma^1_{yy}=2ik u_y^1$. The equation of motion projected along $y$ and the continuity equation then read
\begin{eqnarray}
i\omega \phi^1 &=&-\phi ik u_y^1\;,\\
\rho \phi i\omega u_y^1&=& ik \mathcal{P} \left[-\frac{f'}{f} \phi^1+\frac{2 \left(\mu +g\right)}{\dot \gamma}ik u_y^1  \right]\;.\nonumber
 \end{eqnarray}
In this equation and in the followings, the functions $f$, $f'$, $g$ and $\mu$ are evaluated at the volume fraction $\phi$ of the base state. One finally obtains the dispersion relation
\begin{equation}
\frac{\omega^2}{d^2  |\dot \gamma|^2 k^2}=f' +i \frac{2\left(\mu+g\right)f}{\phi}\;\frac{\omega}{\dot \gamma}\;.
\label{disptrans}
 \end{equation}
When the imaginary term can be neglected in front of the real one, the propagation is found non-dispersive, with a speed
\begin{equation}
c=\frac{\omega}{k}=\sqrt{\frac{1}{\rho}\frac{\partial \mathcal{P}}{\partial \phi}}=|\dot \gamma| d \sqrt{f'}\sim \frac{\sqrt{2} b |\dot \gamma| d}{(\phi_c-\phi)^{3/2}} \;.
\label{cc}
\end{equation}
At a fixed shear rate, the speed of sound is therefore found  to diverge at the jamming point as $(\phi_c-\phi)^{-3/2}$. The imaginary part of the dispersion relation gives the quality factor $\mathcal{Q}$ which, at the jamming point scales as:
\begin{equation}
\mathcal{Q} = - \frac{\phi f'}{2(\mu+g)f}\;\frac{\dot \gamma}{\omega}\sim \frac{1}{(\mu_c+g_c) (\phi_c/\phi-1)}\;\frac{\dot \gamma}{\omega}\;.
\end{equation}
This term diverges  as $(\phi_c-\phi)^{-1}$, which means that the  propagation of this transverse mode becomes asymptotically non-dissipative.

We now consider modes propagating along $z$, of the form $\exp(i\omega t+ikz)$. The perturbation of the strain rate components are $\dot \gamma^1_{xz} = iku^1_x$ and $\dot \gamma^1_{zz} =2ik u^1_z$. The Navier-Stokes equations then read
\begin{eqnarray}
\rho \phi (i\omega u^1_x &+& \dot \gamma u^1_z) = ik \mathcal{P} \left[\left(\mu'+\frac{\mu f'}{f}\right) \phi^1+\frac{2 \mu}{\dot \gamma}ik u^1_x  \right] \;,\nonumber\\
\rho \phi i\omega u^1_z&=& ik \mathcal{P} \left[-\frac{f'}{f}  \phi^1+\frac{2 ( \mu+g )}{\dot \gamma}ik u^1_z -  \frac{2 }{\dot \gamma}ik u^1_x \right] \;, \nonumber\\
\rho i\omega \phi^1 &+& \rho\phi ik u^1_z=0 \;.
\label{Navier}
\end{eqnarray}
This system can be easily solved to derive the dispersion relation, which presents two branches. Approaching the jamming transition, as $\phi \to \phi_c$, the first of these two modes propagates at a speed equal to $c$ (Eq.~\ref{cc}); it presents an elliptical polarization described by
\begin{equation}
\frac{u^1_x}{u^1_z}=-\mu_c+i  \frac{\dot \gamma}{\omega},
\end{equation}
and a quality factor diverging at the jamming point as
\begin{equation}
\mathcal{Q} \sim  \frac{2}{(2\mu_c+g_c) (\phi_c/\phi-1)}\;\frac{\dot \gamma}{\omega}\;.
\end{equation}
The second mode is critically damped and presents a diffusive dispersion relation ($k^2 \propto \omega$) close to jamming:
\begin{equation}
k d \sim \pm \frac{1-i}{\sqrt{2}} \sqrt{ \frac{a \phi_c w}{ b \dot \gamma (\delta \mu)}  (\phi_c-\phi)}.
\end{equation}
It becomes linearly polarized along $x$ as $\phi \to \phi_c$:
\begin{equation}
\frac{u^1_x}{u^1_z} \sim - \frac{i}{(1-\phi/\phi_c)} \;\frac{\dot \gamma}{\omega}\;.
\end{equation}
As it is highly dissipative, this shear mode cannot be observed.

Fig.~\ref{fig:omegastar}a shows that the spontaneous angular frequency coincides, without any adjustable parameter, to the resonant mode $2\pi c/H$ deduced from Eq.~(\ref{cc}). Adding a sinusoidal bulk force on the right hand side of Eqs.~(\ref{Navier}) and applying a no-slip boundary condition at the walls, one obtains the predictions for the resonance curves, with a single adjustable parameter: the second viscosity. As shown in Fig.~\ref{fig:forcing_z}, the resonant frequency, the phase curve and the tails of the amplitude response are largely insensitive to the value of $g$. The fit of the amplitude around the resonance for different volume fractions gives $g(\phi)\simeq g_c \simeq 2$.

{\it Concluding remarks~--} Both transverse and sagittal modes are directly controlled by the dynamical compressibility, which results from the non-affine particle motion  during shearing. Close to the jamming point, the statistical properties of grain trajectories are remarkably insensitive to the nature of the dissipative mechanisms \cite{Andreotti:12,Lerner:12} and to the dynamical regime considered (inertial or overdamped) \cite{Boyer:11,Trulsson:12}. As steric effects dominate, cooperative non-affine motions mostly depend on $\phi$. To move by an average distance $d$ along $x$, a grain actually makes an erratic motion  whose average length diverges as $\sim d (\phi_c-\phi)^{-1}$ \cite{Heussinger:10,Andreotti:12}. By definition of the shear rate, such a displacement occurs at a rate $\sim \dot \gamma$. The dynamical pressure, controlled by inertia, scales as the squared velocity fluctuations and therefore as $\mathcal{P}\sim\rho \dot \gamma^2  d^2 (\phi_c-\phi)^{-2}$ (see Fig.~\ref{fig:omegastar}b). Let us emphasize that this scaling law is different from that obtained in the unsheared case, thermalized \cite{PZ10} or not \cite{Berthier:11}, for which $\mathcal{P} \sim (\phi_c-\phi)^{-1}$ and therefore $c\sim (\phi_c-\phi)^{-1}$.

The propagative modes exhibited in this Letter are therefore specific of the sheared dense liquid granular regime: they neither depend on the restitution coefficient nor on the grain elasticity and the scaling law obeyed by the speed of sound $c$ is a direct consequence of cooperative effects close to jamming. As these modes become asymptotically non-dissipative at the jamming point, they should be observed experimentally and could be used to probe the bulk of granular flows.


{}

\end{document}